\documentclass[aps,prl,twocolumn,floats,showpacs,floatfix,a4paper]{revtex4}

\usepackage{epsfig}
\usepackage{color}
\usepackage{bm}
\usepackage{latexsym}

\begin{document}

\newcommand{\mm}[1]{{\mathbf{#1}}}
\newcommand{\cc}{{\bf\Large C }}
\newcommand{\hide}[1]{}
\newcommand{\tbox}[1]{\mbox{\tiny #1}}
\newcommand{\half}{\mbox{\small $\frac{1}{2}$}}
\newcommand{\sinc}{\mbox{sinc}}
\newcommand{\const}{\mbox{const}}
\newcommand{\tr}{\mbox{tr}}
\newcommand{\intt}{\int\!\!\!\!\int }
\newcommand{\ointt}{\int\!\!\!\!\int\!\!\!\!\!\circ\ }
\newcommand{\eexp}{\mbox{e}^}
\newcommand{\EPS} {\mbox{\LARGE $\epsilon$}}
\newcommand{\ar}{\mathsf r}
\newcommand{\im}{{\cal I}m}
\newcommand{\re}{{\cal R}e}
\newcommand{\bmsf}[1]{\bm{\mathsf{#1}}}
\newcommand{\dd}[1]{\:\mbox{d}#1}
\newcommand{\abs}[1]{\left|#1\right|}
\newcommand{\bra}[1]{\left\langle #1\right|}
\newcommand{\ket}[1]{\left|#1\right\rangle }
\newcommand{\mbf}[1]{{\mathbf #1}}
\newcommand{\eos}{\,.}
\definecolor{red}{rgb}{1,0.0,0.0}

\title{Experimental Study of Active LRC Circuits with ${\cal PT}$- Symmetries}

\author{Joseph Schindler, Ang Li, Mei C. Zheng, F. M. Ellis, Tsampikos Kottos}
\affiliation{Department of Physics, Wesleyan University, Middletown, Connecticut 06459}

\date{\today}

\begin{abstract}
Mutually coupled modes of a pair of  active LRC circuits, one with amplification and another with an equivalent amount 
of attenuation, provide an experimental realization of a wide class of systems where gain/loss mechanisms break the 
Hermiticity while preserving parity-time ${\cal PT}$ symmetry. For a value $\gamma_{\cal PT}$ of the gain/loss strength 
parameter the eigen-frequencies undergo a spontaneous phase transition from real to complex values, while the normal modes 
coalesce acquiring a definite chirality. The consequences of the phase-transition in the spatiotemporal energy evolution 
are also presented.
\end{abstract}

\maketitle

Parity (${\cal P}$) and time -reversal (${\cal T}$) symmetries, as well as their
breaking, belong to the most basic notions in physics. Recently there has been much interest in systems 
which do not obey ${\cal P}$ and ${\cal T}$-symmetries separately but do exhibit a combined ${\cal PT}$
-symmetry. Examples of such ${\cal PT}$-symmetric systems range from quantum field theories and mathematical 
physics \cite{BB98,BBM99,BBJ02} to atomic \cite{GKN08}, solid state \cite{BFKS09,JSBS10} and classical 
optics \cite{MGCM08,L09,GSDMVASC09,RMGCSK10,RKGC10,ZCFK10,L10b,LREKCC11,SXK10}. A ${\cal PT}$-symmetric 
system can be described by a phenomenological "Hamiltonian" ${\cal H}$. Such Hamiltonians may have a real 
energy spectrum, although in general are non-Hermitian. Furthermore, as some parameter $\gamma$ that controls 
the degree of non-Hermiticity of ${\cal H}$ changes, a spontaneous 
${\cal PT}$ symmetry breaking occurs. The transition point $\gamma=\gamma_{\cal PT}$ show the characteristic 
behaviour of an {\it exceptional point} (EP) where both eigenvalues and eigenvectors coallesce (for experimental
studies of EP singularities of lossy systems see Ref.~\cite{Detal01}).
For $\gamma>\gamma_{\cal PT}$, the eigenfunctions of ${\cal H}$ 
cease to be eigenfunctions of the ${\cal PT}$-operator, despite the fact that ${\cal H}$ and the $\mathcal{PT}$-
operator commute \cite{BB98}. This happens because the ${\cal PT}$-operator is anti-linear, and thus the eigenstates 
of ${\cal H}$ may or may not be eigenstates of ${\cal PT}$. As a consequence, in the {\it broken} ${\cal PT}$-
symmetric phase the spectrum becomes partially or completely complex. The other limit where both ${\cal H}$ 
and ${\cal PT}$ share the same set of eigenvectors, corresponds to the so-called {\it exact} $\mathcal{PT}$-symmetric
phase in which the spectrum is real. This result led Bender and colleagues to propose an extension of quantum 
mechanics based on non-Hermitian but ${\cal PT}$-symmetric operators \cite{BB98,BBM99}. The class of non-
Hermitian systems with real spectrum has been extended by Mostafazadeh in order to include Hamiltonians with 
generalized ${\cal PT}$ (antilinear) symmetries \cite{M03}.

While these ideas are still debatable, it was recently suggested that optics can provide a particularly fertile 
ground where ${\cal PT}$-related concepts can be realized \cite{MGCM08} and experimentally investigated 
\cite{GSDMVASC09,RMGCSK10}. In this framework, ${\cal PT}$ symmetry demands that the complex refractive index 
obeys the condition $n({\vec r})=n^*(-{\vec r})$. ${\cal PT}$-synthetic materials can exhibit several intriguing 
features. These include among others, power oscillations and non-reciprocity of light propagation 
\cite{MGCM08,RMGCSK10,ZCFK10}, absorption enhanced transmission \cite{GSDMVASC09}, and unidirectional invisibility 
\cite{LREKCC11}. In the nonlinear domain, such non-reciprocal effects can be used to realize a new 
generation of on-chip isolators and circulators \cite{RKGC10}. Other advances within the framework of ${\cal PT}$-
optics include the study of Bloch oscillations \cite{L09}, the realization of coherent perfect absorbers/lasers 
\cite{L10b} and nonlinear switching structures \cite{SXK10}. Despite all these efforts and the consequent wealth 
of theoretical results associated with ${\cal PT}$-structures, only one experimental realization of a system
with balanced gain and loss, has been reported up to now \cite{RMGCSK10}. These authors studied the light propagation 
in two coupled ${\cal PT}$ symmetric waveguides where the spontaneous ${\cal PT}$-symmetry breaking ``phase transition'' 
\cite{note1}
was confirmed. The analysis has relied on the {\it paraxial approximation} which under appropriate 
conditions maps the 
scalar wave equation to the Schr\"odinger equation, with the axial wavevector playing the role of energy and with a 
{\it fictitious} time, related to the propagation distance along the waveguide axis. However, an experimental 
investigation of ${\cal PT}$-systems in the {\it spatiotemporal} domain has, until now, remained unexplored. 

The purpose of this Letter is to present a simple experimental set-up, which displays all the novel phenomena 
encountered in systems with generalized ${\cal PT}$-symmetries: a pair of coupled oscillators, one with 
gain and the other with loss. This "active" dimer, is implemented with simple 
electronics, and allow a direct observation of a ``phase transition'' from a real 
to a complex eigenfrequency spectrum. At the ${\cal PT}$-breaking point, the normal 
modes coalesce and the relative phase differences of their components acquire a definite value dictated by the 
inductive coupling. We conclude with the investigation of the temporal behavior of the energy. The generic properties 
of pseudo-Hermitian dynamics, are identified and traced back to the properties of the normal modes. Being free of 
theoretical approximations, and due to its relative simplicity in the experimental implementation, the LRC-networks 
with ${\cal PT}$ symmetry, can offer new insights into the study of systems with generalized ${\cal PT}$-symmetries
and a practical means for testing new concepts.

\begin{figure}
\includegraphics[width=1\columnwidth,keepaspectratio,clip]{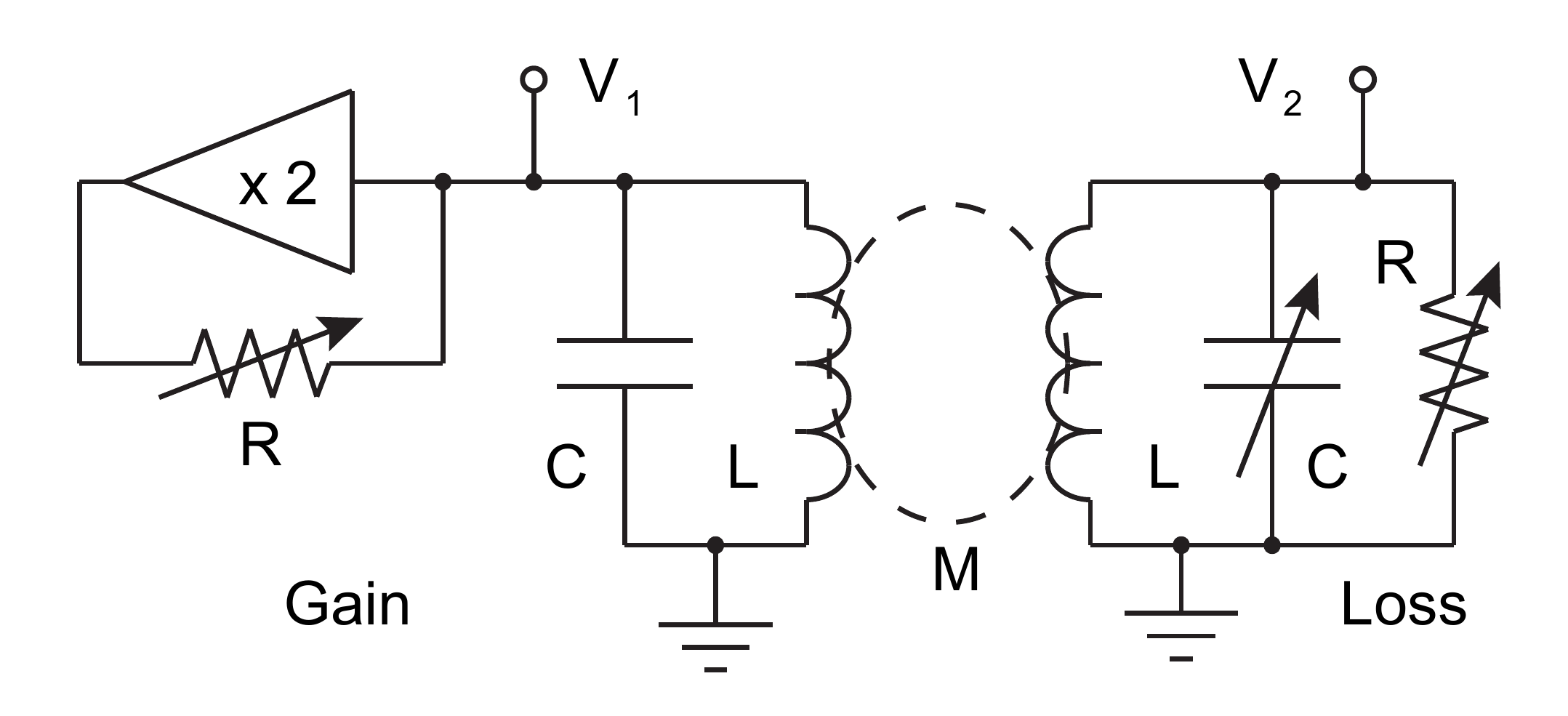}
\caption{Electronic implementation of a ${\cal PT}$-symmetric dimer. The negative resistance gain 
element is provided by feedback from a voltage doubling buffer. The coils are inductively coupled, and $V_1$ and $V_2$ provide access to the system variables.
}
\label{fig:fig1}
\end{figure}


The experimental circuit is shown schematically in Fig.~\ref{fig:fig1}. Each inductor is 
wound with 75 turns of \#28 copper wire on $15 cm$ diameter PVC forms in a $6\times6 mm$ loose bundle for an inductance 
of $L = 2.32 \mu H$. The coils are mounted coaxially with a bundle separation of $4.6 cm$ providing a mutual coupling 
of $\mu = M/L = 0.2$ used for this work. The $C = 10.7 nF$ capacitances are silver-mica, 
in parallel with the $\sim 320 pF$ coil bundle capacitance. The gain side is further trimmed with 
a GR722-M precision condenser. The resistors are carbon composite with the negative resistance 
gain provided by an LM356 op-amp. The isolated natural frequency
of each coil is $\omega_0=1/\sqrt{LC}=2\times10^5$ $s^{-1}$.

The actual experimental circuit deviates from ideal in the following
ways: (1) A resistive component associated with coil wire dissipation is
nulled out with an equivalent gain component applied to each coil; (2) A small
trim is included in the gain buffer for balancing; and (3) Additional LM356 voltage
followers are used to buffer the measured voltages $V_1$ and $V_2$.

It is important to note that the linear nature of the system allows an exact balance
of the ${\cal PT}$ symmetry only to the extent that component imbalance over a time scale
necessary to perform a measurement is negligible. The real system modes
ultimately either exponentially grow to the nonlinearity limit of the buffers, or shrink to zero.
Experimental practice allows only for a marginal determination gain/loss balance. The gain/loss
parameter is set by choosing the loss-side resistance $R$ (in the range  $1-10 k \Omega$ for this work) of Fig.
\ref{fig:fig1} giving $\gamma = R^{-1} \sqrt{L/C}$, and matching the gain-side
$R$ with the help of the gain trim. Our ability to balance the system parameters is estimated
to be approximately $0.1 \%$.

Application of the first and second Kirchhoff's law, for the coupled
circuits of Fig. \ref{fig:fig1}, leads to the set of equations
\begin{eqnarray}
 \label{kirchhoff}
 I_n^C + I_n^R + I_n^L = 0 &;& I_n^R = (-1)^n \gamma \omega_0 Q_n^C \\
 \omega_0^2 Q_1^C = \dot{I}_1^L + \mu \dot{I}_{2}^L &;&
 \omega_0^2 Q_2^C = \dot{I}_2^L + \mu \dot{I}_{1}^L \nonumber
\end{eqnarray}
where $Q$ is the charge, $I$ is the current, and ${\dot I}=dI/dt$. The super-indices $C$, $L$
and $R$ indicate that the quantity is associated with the capacitor, inductor and resistor, while the sub-indices 
correspond to the amplified ($n=1$) and lossy ($n=2$) sides. Simple algebra allows us to re-write Eqs. (\ref{kirchhoff})
for the charges $Q_n^C = CV_n$ in the form:
\begin{eqnarray}
 \label{kirchhoff2}
\frac{d^2Q_1^C}{d\tau^2} &=& -\frac{1}{1-\mu^2} Q_1^C + \frac{\mu}{1-\mu^2} Q_2^C + \gamma \frac{dQ_1^C}{d\tau} \nonumber \\
\frac{d^2Q_2^C}{d\tau^2} &=& \frac{\mu}{1-\mu^2} Q_1^C - \frac{1}{1-\mu^2} Q_2^C - \gamma \frac{dQ_2^C}{d\tau}
\end{eqnarray}
where $\tau\equiv \omega_0 t$. Hence, all frequencies are measured in units of $\omega_0$. Eqs. (\ref{kirchhoff2}) 
are invariant under a combined ${\cal P}$ (i.e. $n=1\leftrightarrow n=2$) and 
${\cal T}$ (i.e. $t\rightarrow -t$) transformation. 

In fact, Eqs. (\ref{kirchhoff2}) can be recasted in a "rate equation" form by making use of a Liouvillian 
formalism
\begin{equation}
\label{liuvilian1}
{d{\bf \Psi}\over d\tau} = {\cal L}  {\bf \Psi};\quad 
 {\cal L} = \left ( \begin{array}{cccc}
  0 & 0 & 1 & 0 \\
  0 & 0 & 0 & 1 \\
 -\frac{1}{1-\mu^2} & \frac{\mu}{1-\mu^2} & \gamma & 0 \\
 \frac{\mu}{1-\mu^2} & -\frac{1}{1-\mu^2} & 0 & -\gamma
 \end{array} \right)
\end{equation}
where ${\bf \Psi}\equiv (Q_1^C, Q_2^C, {\dot Q_1}^C, {\dot Q_2}^C)^T$. It can be shown \cite{Uwe} that there exists 
a similarity transformation mapping the matrix $i{\cal L}$ to a {\cal PT}-symmetric Hamiltonian $H$. This formulation 
opens new exciting directions for applications \cite{Uwe} of generalized ${\cal PT}$-mechanics \cite{M03}.

We are interested in the evolution of 
eigenfrequencies, and normal modes of our system as the gain/loss
parameter $\gamma$ increases. The ${\it exact}$ phase will be associated with the $\gamma$-regime 
for which the eigenfrequencies $\omega_{l}$ are real, while the broken phase corresponds to the regime where 
one or all the eigenfrequencies $\omega_{l}$ become complex.


The eigenfrequencies $\omega_{l}$ can be found either by a direct diagonalization of the 
matrix ${\cal L}$, or by solving the secular equation resulting from Eq. (\ref{kirchhoff2}) after the substitution $Q_n^C=
A_n \exp(i\omega \tau)$. We get:
\begin{eqnarray}
\label{frequencies}
 \omega_{1,4} &= \pm \sqrt{-\frac{2+\gamma^2(\mu^2-1)+\sqrt{4(\mu^2-1)
 +(2+\gamma^2(\mu^2-1))^2}}{2(\mu^2-1)}} \\ 
 \omega_{2,3} &= \pm \sqrt{\frac{-2+\gamma^2(\mu^2-1)+\sqrt{4(\mu^2-1)
 +(2+\gamma^2(\mu^2-1))^2}}{2(\mu^2-1)}}\nonumber
\end{eqnarray}
For $\gamma=0$, we have two frequency pairs $\omega_{1,4}=\pm \sqrt{1/(1-\mu)}$ and $\omega_{2,3}=\pm 
\sqrt{1/(\mu+1)}$. These modes are associated with the pair of double degenerate frequencies $\omega=
\pm 1$ related to a single isolated circuit ($\mu=0$). At $\gamma=\gamma_{\cal PT}$, these eigenmodes 
undergo a level crossing and branch out into the complex plane, with
\begin{equation}
\gamma_{\cal PT} = 1/\sqrt{1-\mu} - 1/\sqrt{1+\mu}
\label{gammaPT}
\end{equation}
Near $\gamma_{\cal PT}$ the eigenvalues display the characteristic behavior $|\omega|\propto\pm \sqrt{
\gamma^{2}- \gamma_{\cal PT}^{2}}$. This square root singularity is a generic feature of the ${\cal PT}$
-symmetry breaking. A second crossing between the pairs of degenerate frequencies (and 
a new branching) happens for a larger value $\gamma_{2}=\sqrt{1/(1-\mu)} +\sqrt{1/(1+\mu)}$. 
For $\gamma>\gamma_{2}$ all frequencies are imaginary. Since this large $\gamma$ regime is 
physically impractical, we will confine ourselves to values of $\gamma<\gamma_{2}$.

\begin{figure}
\includegraphics[width=1\columnwidth,keepaspectratio,clip]{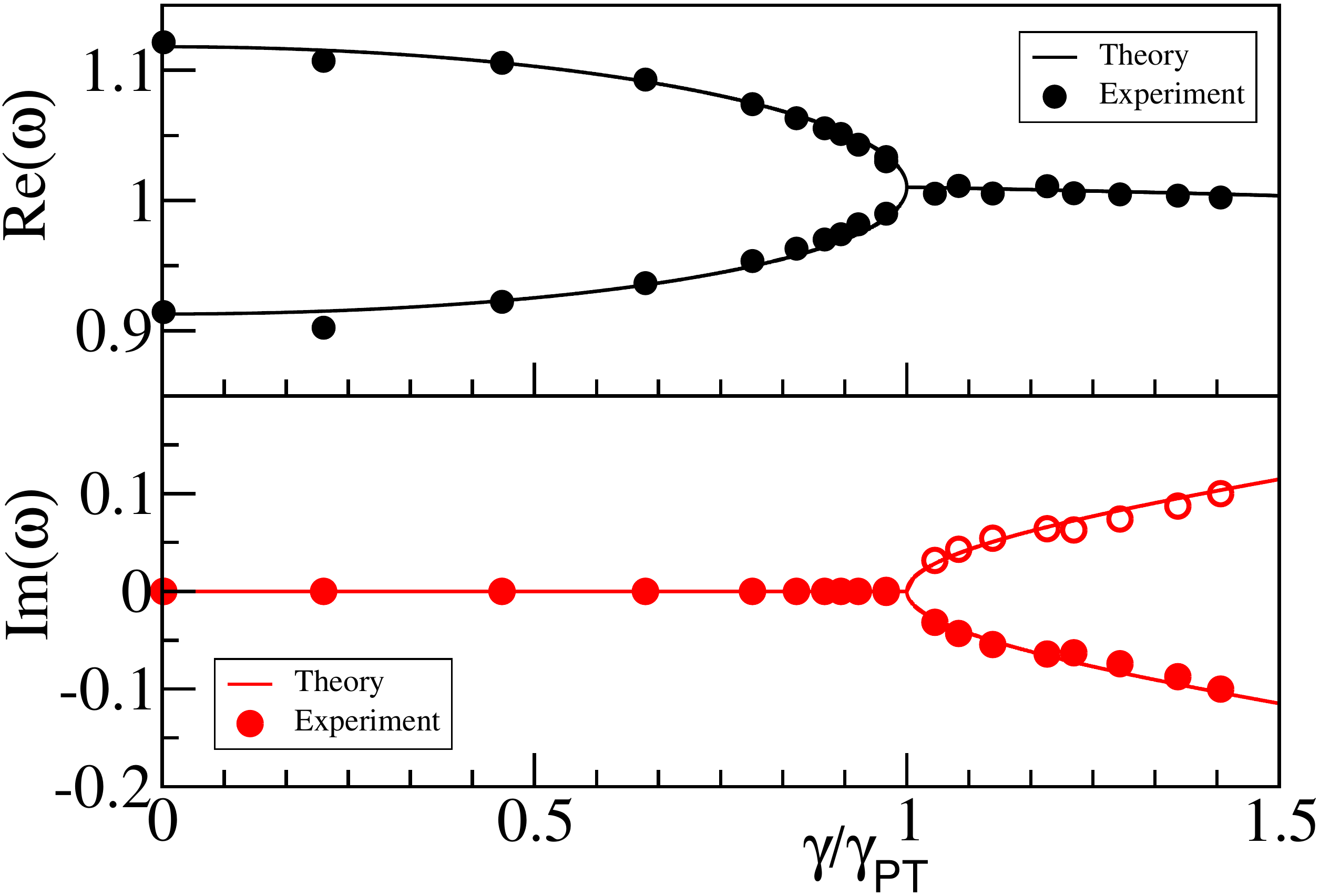}
\caption{Parametric evolution of the experimentally measured eigenfrequencies, vs. the normalized gain/
loss parameter $\gamma/\gamma_{\cal PT}$. A comparison with the theoretical results of Eq. (\ref{frequencies}),
indicate an excellent agreement. In all cases, we show only the ${\cal R}e (\omega_{l})>0$ eigenfrequencies. 
The open circles in the lower panel are reflections of the experimental
 data (lower curve) with respect to the ${\cal I}m(\omega)=0$ axis.
}
\label{fig:fig2}
\end{figure}

In Fig. \ref{fig:fig2} we report our measurements for the frequencies and compare them with Eq. 
(\ref{frequencies}). Our set-up allows detailed analysis for gain/loss parameters $\gamma$ on either 
side of the ${\cal PT}$-phase transition point. In the exact phase, $\gamma < \gamma_{\cal PT}$, the 
eigenfrequencies are obtained by trimming both the gain and capacitance balance until both modes 
become simultaneously marginal. The imaginary part of the frequency is then zero by construction. Individual 
modes are then measured by slightly unbalancing the trim for marginal oscillation of that mode. The amplitude 
grows to the nonlinearity limited amplitude, and the voltage waveforms in both coils are captured and 
analyzed for frequency and relative phase. The values obtained are checked to assure that they are independent 
of the small imbalances applied, typically less than a 0.3\%.

In the broken phase, $\gamma > \gamma_{\cal PT}$, marginal oscillation is not possible:
the gain and capacitance trim are kept fixed at the values obtained in the exact
phase near $\gamma_{\cal PT}$. The resistances $R$ (and $-R$) are inserted, and the gain
side coil is temporarily short-circuited to prevent oscillation. The short is then
removed, and the subsequent oscillatory growth triggers the waveform capture. The
waveforms in both coils are analyzed including an exponential growth factor. With this method, only
the exponentially growing mode with ${\cal I}m(\omega) < 0$ is observed.  

Very close to $\gamma \sim \gamma_{\cal PT}$ attempts to trim the
dimer to the marginal configuration result in either $V=0$ (too small gain),
or a chaotic interplay of the two modes with the op-amp nonlinearity if the gain is
larger. This behavior serves as an indication that $\gamma_{\cal PT}$ has been exceeded.

\begin{figure}
\includegraphics[width=0.85\columnwidth,keepaspectratio,clip]{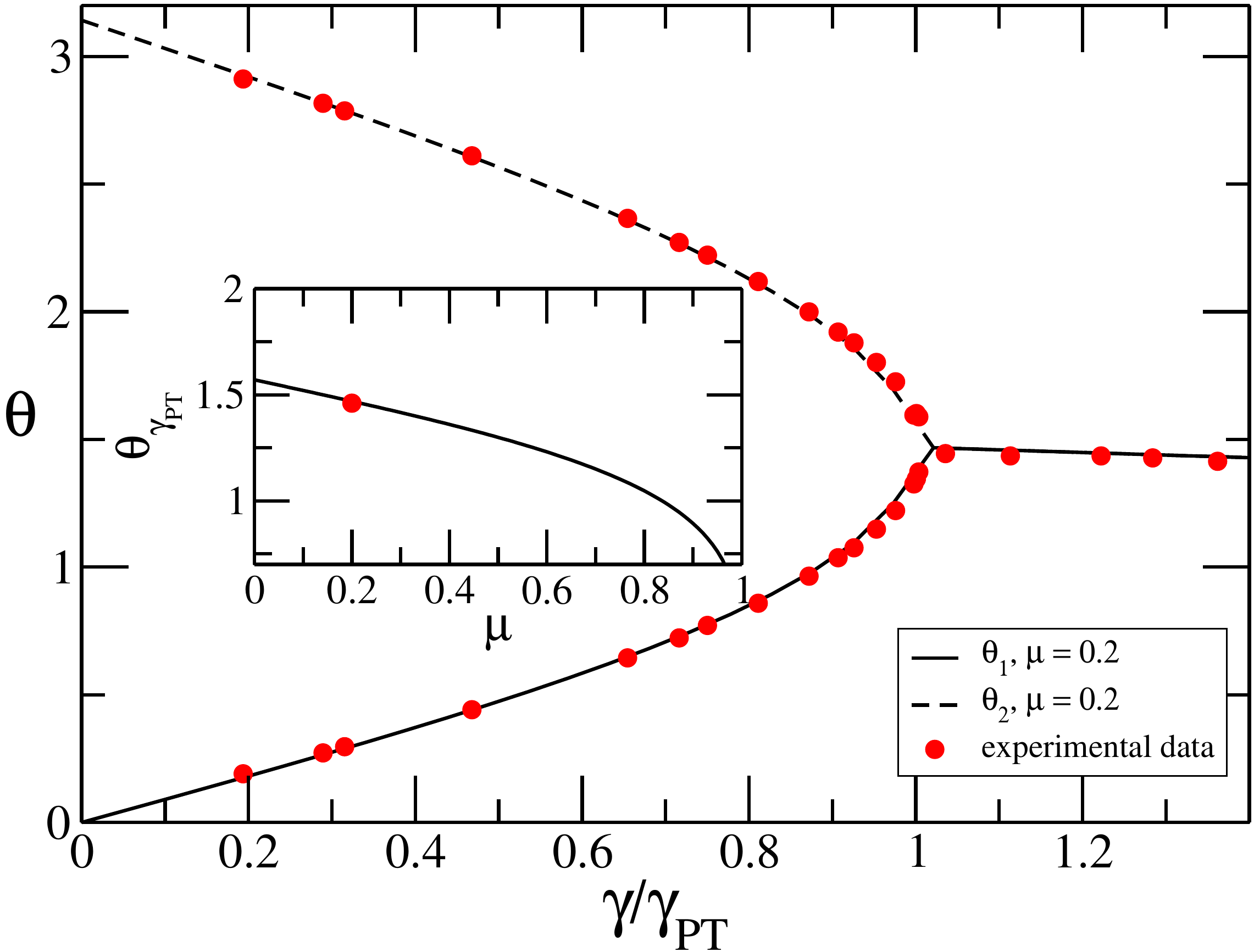}
\caption{Parametric evolution of the phase difference $\theta_{l}=\phi_2^{(l)}-\phi_1^{(l)}$. The experimental 
measurements for $\mu=0.2$ are shown in filled circles and nicely match the theoretical predictions (lines).
The theoretical $\theta_{\cal PT}(\mu)$ is shown in the inset.
}
\label{fig:fig3}
\end{figure}

Another manifestation of ${\cal PT}$-symmetry is the relative phase difference $\theta_{l}=\phi_2^{(l)}-
\phi_1^{(l)}$ between the two components $(Q_1^{(l)}, Q_2^{(l)})^T = (\left|Q_1^{(l)}\right|\exp(i\phi_1^{(l)}), 
\left|Q_2^{(l)}\right| \exp(i\phi_2^{(l)}))^T$ of the $l-$th eigenmode. Experimentally, $Q_n^{(l)}(t) = CV_n(t)$ 
when mode $(l)$ is marginally oscillated. The ${\cal PT}$ symmetry imposes the condition that the 
magnitude of the two components of the charge vector are equal to one-another in the exact phase. For 
$\gamma=0$, the phases corresponding to the symmetric and antisymmetric combination are $\theta_{1}=0$ and 
$\theta_{2}=\pi$, respectively. When $\gamma$ is subsequently increased and the system is below the ${\cal PT}$ 
threshold, the eigenstates are not orthogonal and their phases can be anywhere (depending on $\gamma/
\gamma_{\cal PT}$) in the interval $[0,\pi]$. An example of the parametric evolution of phases is reported in 
Fig. \ref{fig:fig3} where the experimental measurements are plotted together with the theoretical results.
The value of phase difference $\theta_{\cal PT}(\mu)\equiv \theta(\mu, \gamma_{\cal PT})$ at $\gamma=\gamma_{\cal PT}$ 
can be calculated analytically and it is given by the expression:
\begin{equation}
\label{phicr}
\theta_{\cal PT}(\mu)=
\arccos\left(\sqrt{1-\sqrt{1-\mu^2}}/\sqrt{1+\sqrt{1+\mu^2}}\right)
\end{equation}
We note that in the limit of $\mu\rightarrow 0$ we get $\theta_{\cal PT}=\pi/2$, corresponding to a "circular" 
polarization of the eigenmode. The opposite limit of $\mu\rightarrow 1$ is associated with a broken time-reversal 
symmetry \cite{DHKMRS11} and is reflected in the strong asymmetry of the ${\cal L}$-matrix for large 
$\mu$-values.

The signatures of ${\cal PT}$-symmetry and the transition from the exact phase to the
broken phase are reflected in the temporal behavior of our system. We have traced these universal features
by studying the time-dependence of the total capacitance energy:
\begin{equation}
\label{Ectot}
E_C^{\rm tot}(\tau)= {1\over 2C}(Q_1(\tau),Q_2(\tau)) (Q_1(\tau),Q_2(\tau))^T.
\end{equation}
The initial condition used in the experiment corresponds to the case $I_1^{L} =I_{\rm init}$ with all other 
dynamical variables zero. With the 
appropriate $R$ inserted and the dimer trim either marginal (exact phase) or fixed (broken phase), the 
initial current is injected into the gain side by connecting the $V_1$ node of Fig. \ref{fig:fig1} to 
a voltage source through a resistor. Oscillation in the broken phase is suppressed by the additional AC 
dissipation. Again, waveform capture is triggered by removing the injection resistor.

\begin{figure}
\includegraphics[width=1\columnwidth,keepaspectratio,clip]{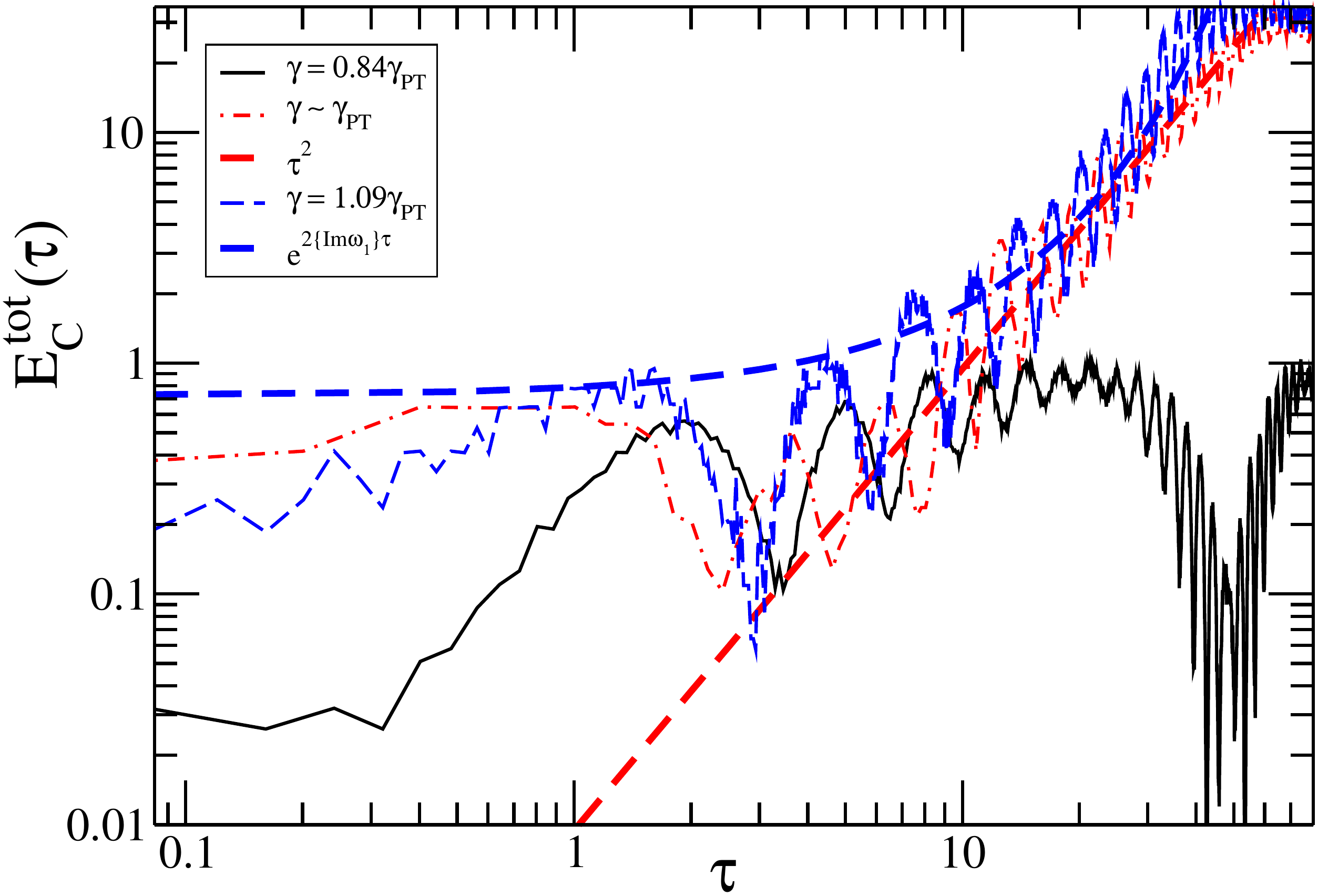}
\caption{Experimentally measured temporal dynamics of the capacitance energy $E_C^{\rm tot}(\tau)$ of the 
total system for various $\gamma$-values. As $\gamma\rightarrow \gamma_{\cal PT}$ the $\tau^2$ behavior 
signaling the spontaneous ${\cal PT}$-symmetry breaking is observed.
}
\label{fig:fig4}
\end{figure}

Even though the frequencies are real for $\gamma< \gamma_{\cal PT}$, the total energy of the system $E^{\rm tot}$ 
is not conserved. Instead, we expect power oscillations which are due to the unfolding of the non-orthogonal 
eigenmodes \cite{BB98,MGCM08,ZCFK10,RMGCSK10}. This is evident in the temporal behavior of $E_C^{\rm tot}(\tau)$ (see Fig. 
\ref{fig:fig4}). For $\gamma> \gamma_{\cal PT}$ the dynamics is unstable and $E_C^{\rm tot}(\tau)$ grows 
exponentially with a rate given by the maximum imaginary eigenvalue $\max\{{\cal I}m(\omega_{l})\}$
(see Fig. \ref{fig:fig4}).

The most interesting behavior appears at the spontaneous ${\cal PT}$-symmetry breaking point $\gamma=\gamma_{
\cal PT}$. At this point the matrix ${\cal L}$ has a defective eigenvalue. In this case, the evolution $U=\exp({\cal L}\tau)$
can be calculated from the Jordan decomposition of ${\cal L}$ as ${\cal J}=S{\cal L}S^{-1}$. Having in mind the
form of the exponential of a Jordan matrix, it follows immediately that linear growing terms appear in the evolution
of the charge vector $\left(Q_1(\tau), Q_2(\tau)\right)^T$ \cite{H10}. This results in a quadratic increase of the 
capacitance energy i.e. $E_C^{\rm tot}(\tau)\sim \tau^2$. 
Although all systems typically becomes very sensitive to parameters near a critical
point, we are able to control the circuit elements sufficiently well to observe the
approach to the predicted $\tau^2$ behavior of the energy. The time range is limited
by the dynamic range of our circuit linearity.

In conclusion, we demonstrated that a pair of coupled, active LRC circuits, one with amplification and the other 
with equivalent attenuation, exhibits ${\cal PT}$ symmetry. This minimal example, which is experimentally simple and 
mathematically transparent, displays all the universal phenomena encountered in systems with generalized 
${\cal PT}$-symmetries. At 
the same time the accessibility of experimental quantities of interest enable us to perform accurate investigations 
and comparisons with theoretical predictions. We envision that their use will allow experimental studies in the 
spatiotemporal domain of more complicated structures, and shed new light on novel scattering phenomena  \cite{Joye} 
recently proposed in the realm of generalized ${\cal PT}$-symmetries \cite{L10b,LREKCC11,CDV07,M09,S10}. From 
integrated tuning of antenna arrays to real-time control of exotic meta-materials, the wealth of problems which have 
their counterparts in ${\cal PT}$ electronic circuits, and the lessons they can teach, are far from being exhausted.

\begin{acknowledgments}
We thank D. Christodoulides, B. Dietz, V. Kovanis, and A. Richter for useful discussions. This research was supported 
by an AFOSR No. FA 9550-10-1-0433 grant and by an NSF ECCS-1128571 grant. 
\end{acknowledgments}



\begin{thebibliography}{99}
\bibitem{note1} We use the term ``phase transition" not in the standard thermodynamic sense but
rather in the frame of ${\cal PT}$-literature (see for example \cite{BB98}).
\end{thebibliography}
\end{document}